\documentclass[11pt]{article}
\usepackage{spconf,amsmath,epsfig}
\usepackage{spconf,graphicx}
\usepackage{subfigure}
\usepackage{amssymb,amsmath}
\usepackage[portuguese]{babel}
\usepackage[utf8]{inputenc}
\usepackage{bm,bbm}
\usepackage{booktabs}
\usepackage{multirow}
\usepackage{rotating}
\usepackage{microtype}

\title{A NEW ALGORITHM OF SPECKLE FILTERING USING STOCHASTIC DISTANCES}
\name{
Leonardo Torres, Tamer Cavalcante, Alejandro C. Frery
}
\address{
Universidade Federal de Alagoas -- UFAL\\
Laborat\'orio de Computa\c c\~ao Cient\'ifica e An\'alise Num\'erica -- LaCCAN\\
57072-970, Macei\'o, AL -- Brazil
}

\usepackage{setspace}

\begin{document}
%
\maketitle
\begin{abstract}
This paper presents a new approach for filter design based on stochastic distances and tests between distributions.
A window is defined around each pixel, overlapping samples are compared and only those which pass a goodness-of-fit test are used to compute the filtered value.
The technique is applied to intensity SAR data with homogeneous regions using the Gamma model.
The proposal is compared with the Lee's filter using a protocol based on Monte Carlo.
Among the criteria used to quantify the quality of filters, we employ the equivalent number of looks, line and edge preservation.
Moreover, we also assessed the filters by the Universal Image Quality Index and the Pearson's correlation on edges regions.
\end{abstract}
\section{Introduction}\label{sec:intro}

Synthetic Aperture Radar (SAR) data are generated by a system of coherent illumination and are affected by the interference coherent of the signal.
It is known that these data incorporate a granular noise that degrades its quality, known as speckle noise, which is also present in the laser, ultrasound-B, and sonar imagery~\cite{Goodman1976}.
The noise makes the segmentation, extraction, analysis and, classification of objects and information in the image hard tasks.

Statistical analysis is essential for dealing with speckled data.
It provides comprehensive support for developing procedures for interpreting the data efficiently, and to simulate plausible images~\cite{Gao2010}.
In this paper, the multiplicative model was used to describe the speckle noise (see Section~\ref{sec:model}).

Different statistical distributions are proposed in the literature to describe speckle data.
In this paper we use the Gamma distribution to describe the speckle noise, and a constant to characterize the ground truth~\cite{Gao2010}.
The proposed filter is a local nonlinear procedure.
It is based on stochastic distances between distributions, as presented in~\cite{Nascimento2010}.

The paper is organized as follows:
Section~\ref{sec:model} presents the statistical modeling used to describe speckle data.
Section~\ref{sec:distances} describes the new method for filtering speckle.
Section~\ref{sec:assessment} presents the metrics for assessing the quality of the filtered images.
Sections~\ref{sec:results} and~\ref{sec:conclu} present the results and conclusions.

\section{The Multiplicative Model}\label{sec:model}

According to reference~\cite{Goodman1976}, the multiplicative model can be used to describe SAR data.
This model asserts that the intensity observed in each pixel is the outcome of the random variable 
$Z\colon\Omega \rightarrow \mathbbm{R}_+$ which, in turn, is the product of two independent random variables: 
$X\colon\Omega \rightarrow \mathbbm{R}_+$, that characterizes the backscatter; and
$Y\colon\Omega \rightarrow \mathbbm{R}_+$, which defines the intensity of the speckle noise.
The distribution related to the observed intensity $Z=XY$ is completely specified by the distributions proposed for $X$ and $Y$.

This paper focus is homogeneous regions in intensity images, so the constant $X\thicksim\lambda>0$ defines the backscatter, and $Y\thicksim \Gamma(L,L)$ models the speckle noise by a Gamma distribution (with expected value $\mathbbm{E}(Y)=1$), where $L$ is equivalent number of looks.
Thus, it follows that $Z\thicksim \Gamma\left(L,{L}/{\lambda}\right)$ and its density is
\begin{equation}
f_Z(z;L,\lambda) = \frac{L^L}{\lambda^L\Gamma(L)} z^{L-1} \exp\Big\{ \frac{-Lz}{\lambda} \Big\}, 
\label{eq:densgamma}
\end{equation}
$L\geq1, z,\lambda > 0$.

\section{Stochastic Distances Filter}\label{sec:distances}

The proposed filter is local and nonlinear.
It is based stochastic distances and tests between distributions~\cite{Nascimento2010}, obtained from the class of ($h,\phi$)-divergences.
The proposal employs the neighborhoods defined by Nagao and Matsuyama~\cite{NagaoMatsuyama}.

Each filtered pixel has a $5\times5$ neighborhood, within which nine areas are defined and treated as different samples.
Denote $\bm{\widehat{\theta}_1}$ the estimated parameter in the central $3\times3$ neighborhood,
and $\big(\bm{\widehat{\theta}}_2,\ldots,\bm{\widehat{\theta}}_{9}\big)$ the estimated parameters in the eight remaining areas.
To account for possible departures from the homogeneous model, we estimate $\bm{\widehat{\theta}}_i=(L_i,\lambda_i)$, $i=\{1,\dots,9\}$ by maximum likelihood.

The proposal is based on the use of stochastic distances on small areas within the filtering window.
Consider $Z_1$ and $Z_i$ random variables defined on the same probability space, characterized by the densities $f_{Z_1}(z_1;\bm{\theta}_1)$ and $f_{Z_i}(z_i;\bm{\theta}_i)$, respectively, where $\bm{\theta_1}$ and $\bm{\theta_i}$ are parameters.
Assuming that both densities have the same support $I \subset \mathbbm{R}$, the ($h,\phi$)-divergence between $f_{Z_1}$ and $f_{Z_i}$ is given by
\begin{equation}
D_{\phi}^{h}(Z_1,Z_i) = h \Big( \int_{x\in I}\;\phi \Big( \frac{f_{Z_1}(x;\bm{\theta}_1)}{f_{Z_i}(x;\bm{\theta}_i)} \Big) \;f_{Z_i}(x;\bm{\theta}_i)\;\mathrm{d}x \Big),
\end{equation}
where $h\colon (0,\infty)\rightarrow[0,\infty)$ is a strictly increasing function with $h(0)=0$ and $h'(x)>0$, $\phi\colon (0,\infty)\rightarrow[0,\infty)$ is a convex function for all $x \in \mathbbm{R}$.
Choices of the functions $h$ and $\phi$ result in several divergences.

Divergences sometimes do not obey the requirements to be considered distances.
A simple solution, described in~\cite{Nascimento2010}, is to define a new measure $d_{\phi}^{h}$ given by
\begin{equation}
d^{h}_{\phi}(\bm{\widehat{\theta}}_1,\bm{\widehat{\theta}}_i) = \frac{{D_{\phi}^{h}(Z_1,Z_i)+D_{\phi}^{h}(Z_i,Z_1)}}{2}.
\end{equation}
Distances, in turn, can be conveniently scaled in order to present good statistical properties that make them test statistics~\cite{Nascimento2010}:
\begin{equation}
S_{\phi}^{h}(\bm{\widehat{\theta}}_1,\bm{\widehat{\theta}}_i)=\frac{2mnk}{m+n}\;d^{h}_{\phi}(\bm{\widehat{\theta}}_1,\bm{\widehat{\theta}}_i),
\end{equation}
where $\bm{\widehat{\theta}}_1$ e $\bm{\widehat{\theta}}_i$ are maximum likelihood estimators based on samples size $m$ and $n$, respectively, and $k=\big(h'(0)\phi''(1)\big) ^{-1}$.
The null hypothesis $\bm{\theta_1}=\bm{\theta_i}$ is rejected at a level $\eta$, if $\Pr(S_{\phi}^{h}>\eta)$, and since under mild conditions $S_{\phi}^{h}$ is $\chi^2_M$ asymptotically distributed, being $M$ the dimension of $\bm{\theta_1}$, the test is well defined.
Details can be seen in the work by Salicr\'u et al.~\cite{Salicru1994}. 
The statistical test derived in this paper was the Kullback-Leibler test:
\begin{equation}
S_{K\!L} = \frac{2mn}{m+n}\;\widehat{L}\bigg(\frac{\widehat{\lambda}_1^2+\widehat{\lambda}_i^2}{2\widehat{\lambda}_1 \widehat{\lambda}_i}-1\bigg).
\end{equation}

The filtering procedure consists in checking which regions can be considered as coming from the same distribution that produced the data which comprises the central block.
The sets which are not rejected are used to compute a local mean.
If all the sets are rejected, the filtered value is updated with the average on the $3\times3$ neighborhood around the filtered pixel.

\section{Image Quality Assessment}\label{sec:assessment}

\begin{figure*}[hbt]
 \centering
 \subfigure[Phantom]{\label{fig:phantom}
 \includegraphics[width=.24\linewidth]{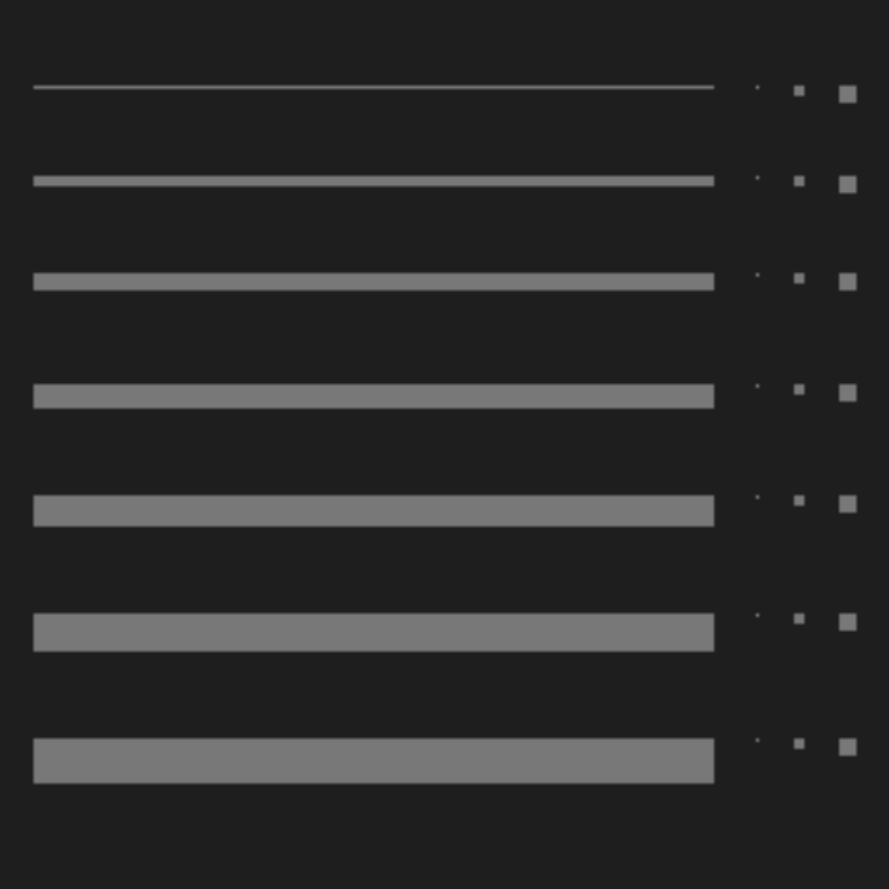}}
 \subfigure[Corrupted, $4$-looks]{\label{fig:corrupt_4looks}
 \includegraphics[width=.24\linewidth]{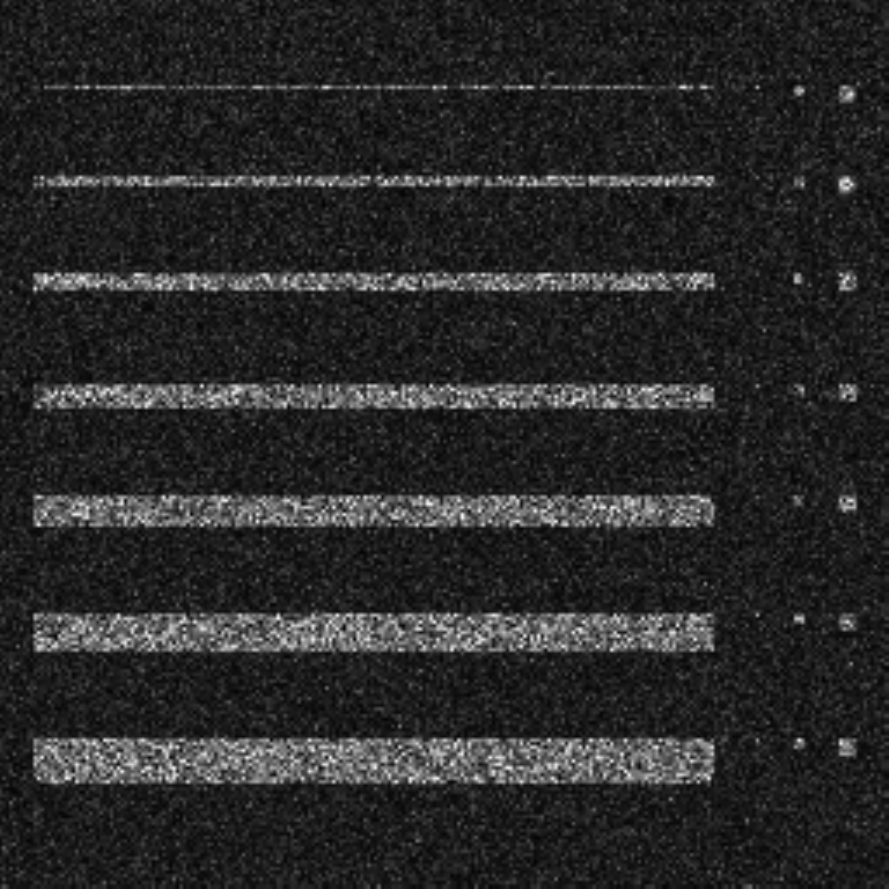}}
 \subfigure[Lee filter]{\label{fig:LeeFilter_4looks}
 \includegraphics[width=.24\linewidth]{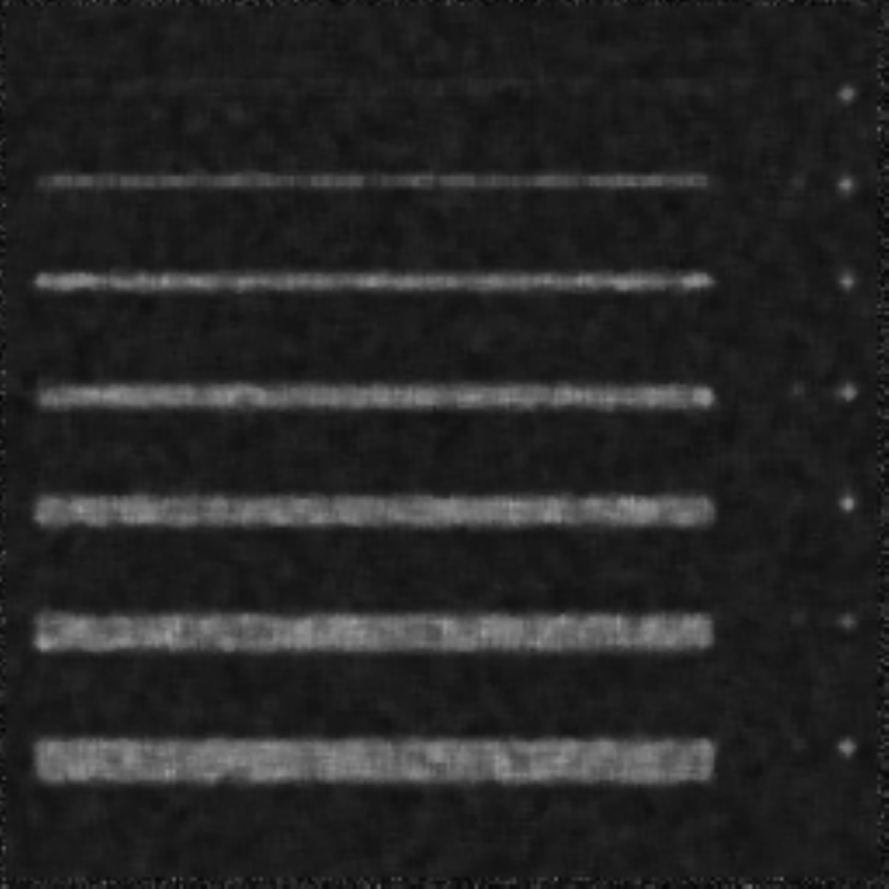}}
 \subfigure[Kullback-Leibler filter]{\label{fig:KL_Filter_4looks}
 \includegraphics[width=.24\linewidth]{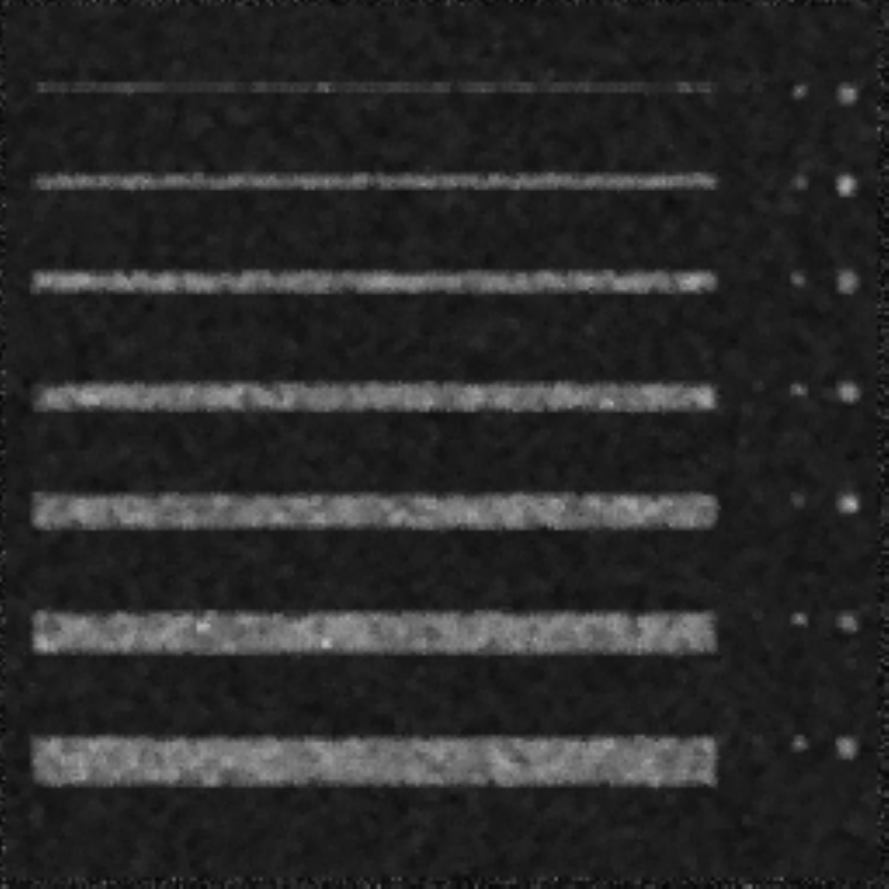}}
 \caption{Lee’s Protocol phantom, speckled data and filtered images.}
 \label{fig:protocol}
\end{figure*}

Image quality assessment in general, and filter performance evaluation in particular, are hard tasks~\cite{Moschetti2006,UIQIndex}.
Moschetti et al~\cite{Moschetti2006} discussed the need of making a Monte Carlo study when assessing the performance of image filters.
They proposed a protocol which consists of using a phantom image (see Figure~\ref{fig:phantom}) corrupted by speckle noise (see Figure~\ref{fig:corrupt_4looks}). 
The experiment consists of simulating corrupted images as matrices of independent samples of some distribution with different parameters.
Every simulated image is subjected to filters, and the results are compared (see Figures~\ref{fig:LeeFilter_4looks} and~\ref{fig:KL_Filter_4looks}).
Among the criteria used to quantify the quality of the filters, we employ the equivalent number of looks (\textsf{NEL}), line preservation and edge preservation.
A ``good'' technique must combat speckle and, at the same time, preserve details as well as relevant information.

We also assessed the filters by the universal image quality index \cite{UIQIndex} and the correlation measure $\beta_{\rho}$.
The universal image quality index is defined by
\begin{equation}
Q = \frac{s_{xy}}{s_x s_y} \frac{2\overline{xy}}{\overline{x}^2 + \overline{y}^2} \frac{2 s_x s_y}{s_x^2 + s_y^2},
\end{equation}
where $s_\bullet^2$ and $\overline{\bullet}$ denote the sample variance and mean, respectively.
The range of $Q$ is $[-1,1]$, being $1$ the best value.
The quantity
\begin{equation}
\beta_{\rho} = \frac{\sum_{j=1}^{n} (x_j-\bar{x})(y_j-\bar{y})}{\sqrt{\sum_{j=1}^{n} (x_j-\bar{x})^2 \sum_{j=1}^{n} (y_j-\bar{y})^2}},
\end{equation}
is a correlation measure is between the Laplacians of images $X$ and $Y$, where $\bullet_j$ and $\overline{\bullet}$ denote the gradient values of the $jth$ pixel and mean of the images $\nabla^2 X$ and $\nabla^2 Y$, respectively. 
The range of $\beta_{\rho}$ is $[-1,1]$, being $1$ perfect correlation.

\section{Results and Analysis}\label{sec:results}

\begin{figure*}[hbt]
 \centering
 \subfigure[Equivalent Number of Looks]{\includegraphics[trim=0 0 0 50, width=.32\linewidth]{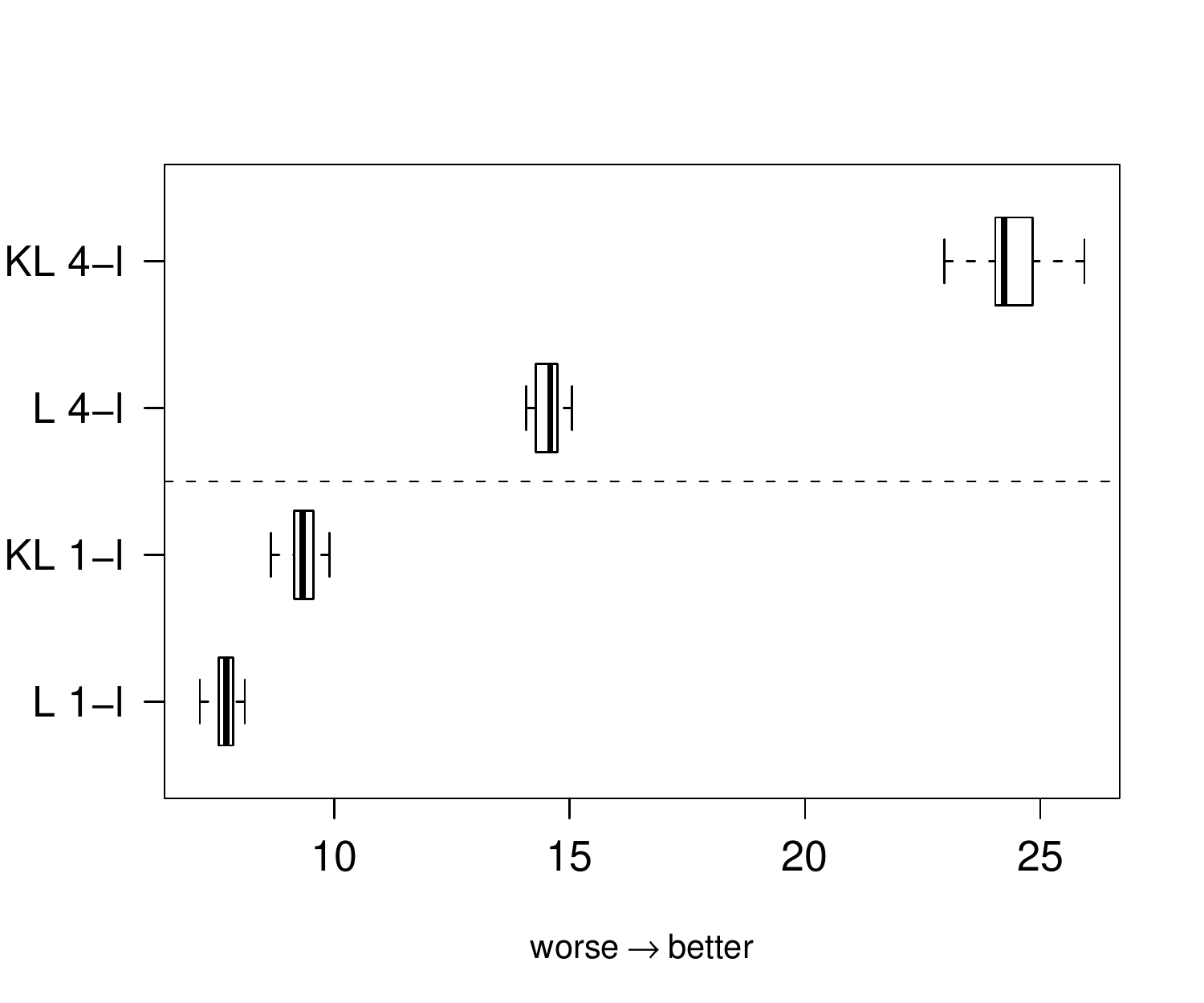}}
 \subfigure[Line Preservation]{\includegraphics[trim=0 0 0 50, width=.32\linewidth]{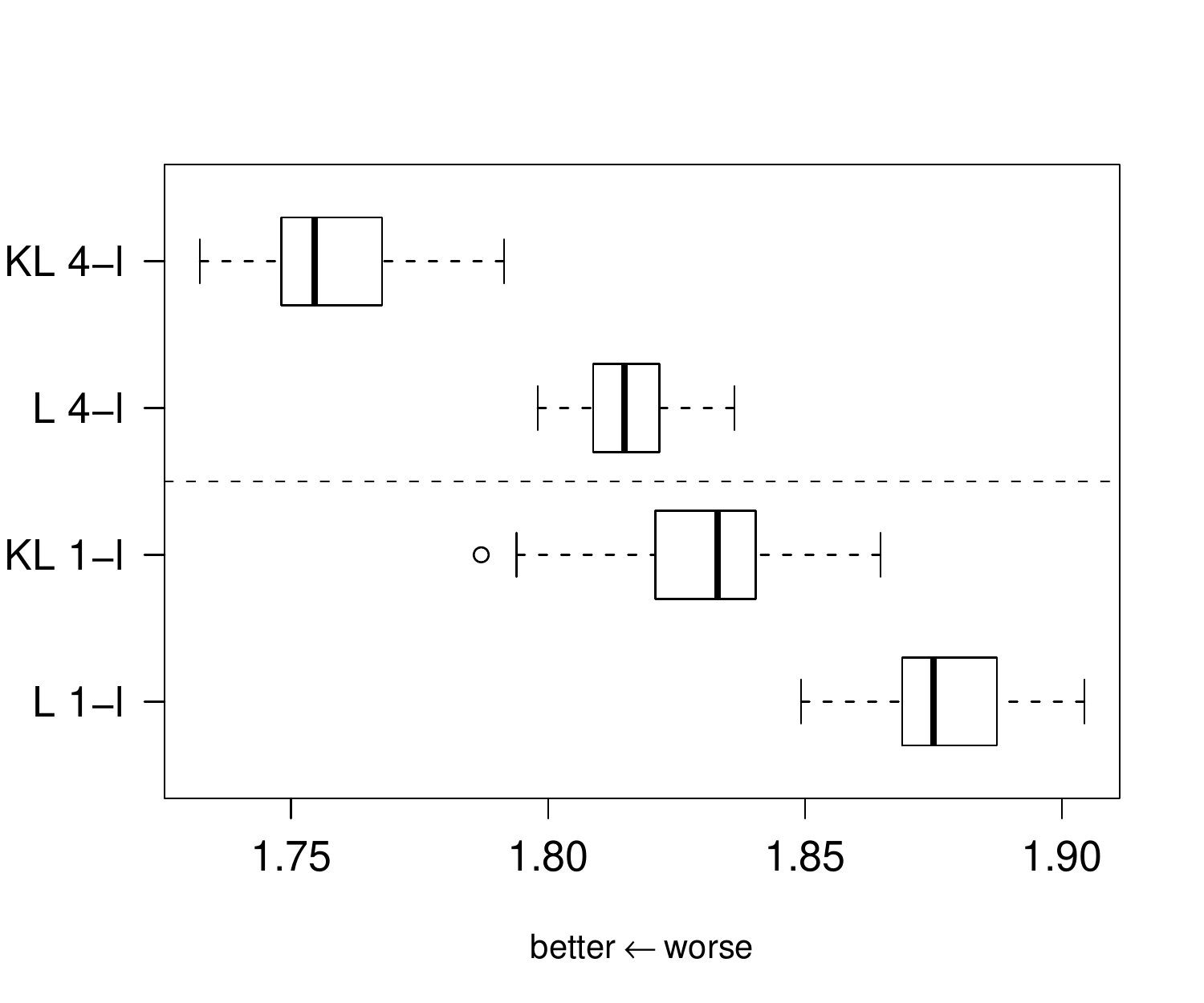}}
 \subfigure[Edge Gradient]{\includegraphics[trim=0 0 0 50, width=.32\linewidth]{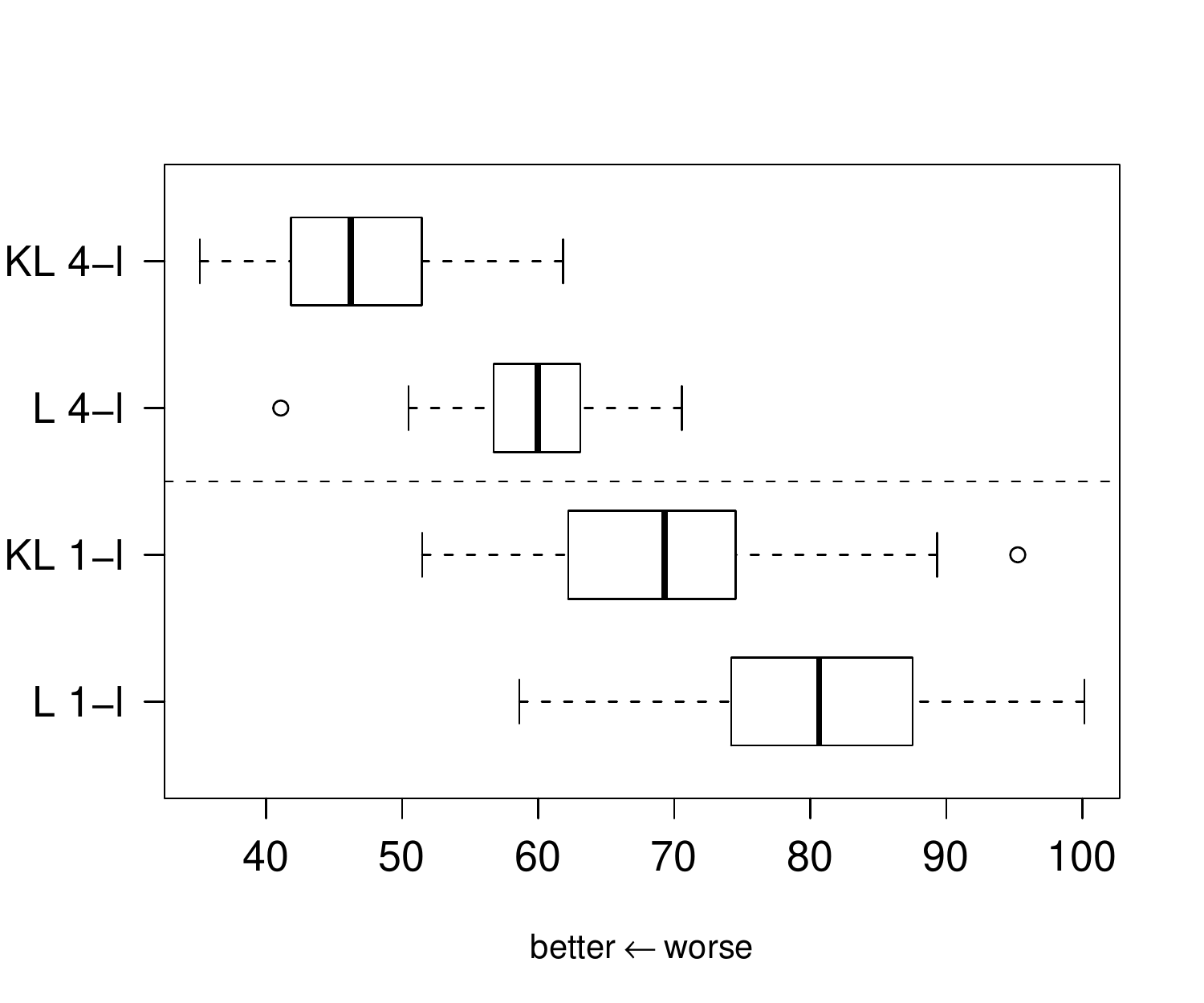}}
 \subfigure[Edge Variance]{\includegraphics[trim=0 0 0 50, width=.32\linewidth]{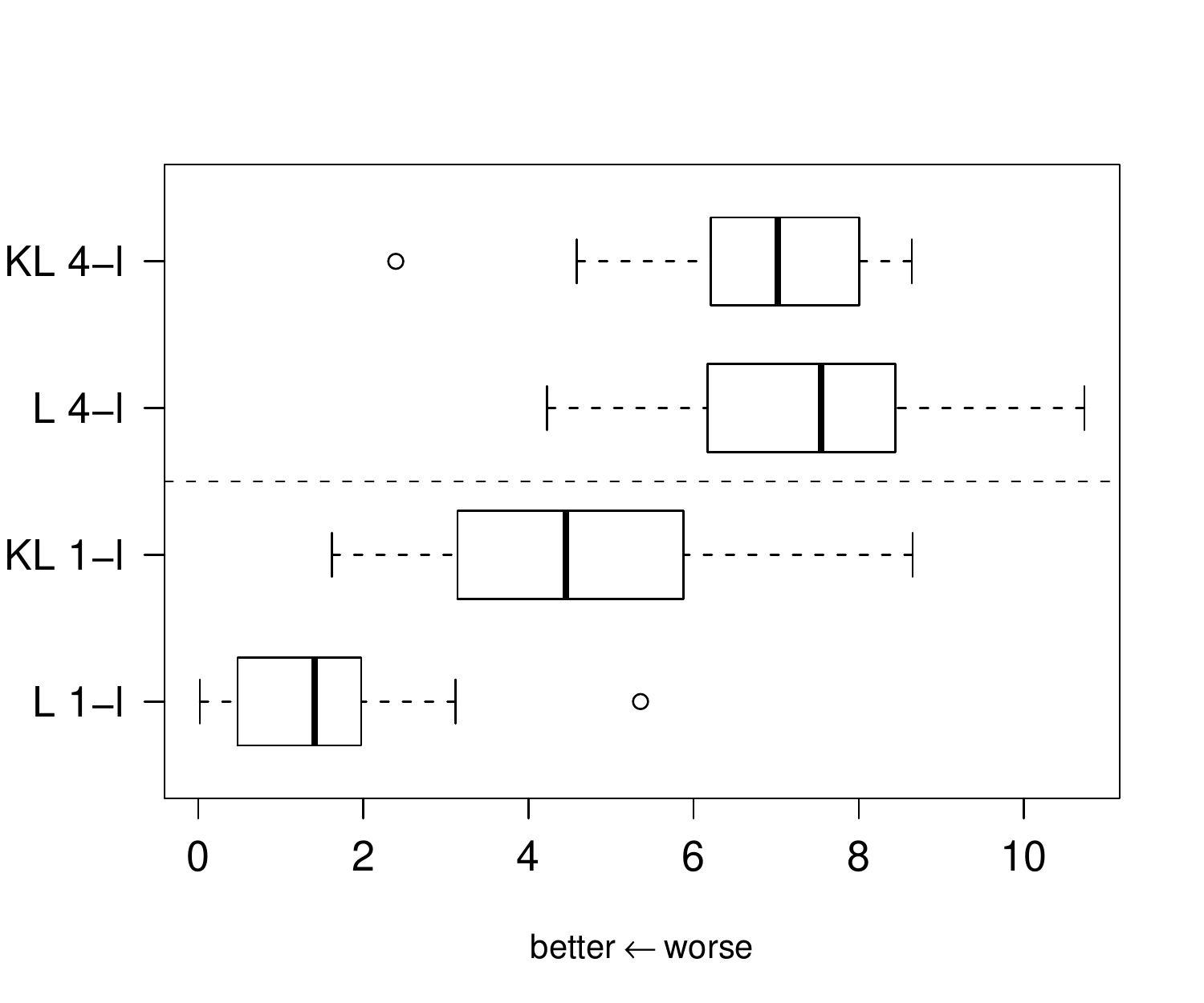}}
 \subfigure[Values of $Q$]{\includegraphics[trim=0 0 0 50, width=.32\linewidth]{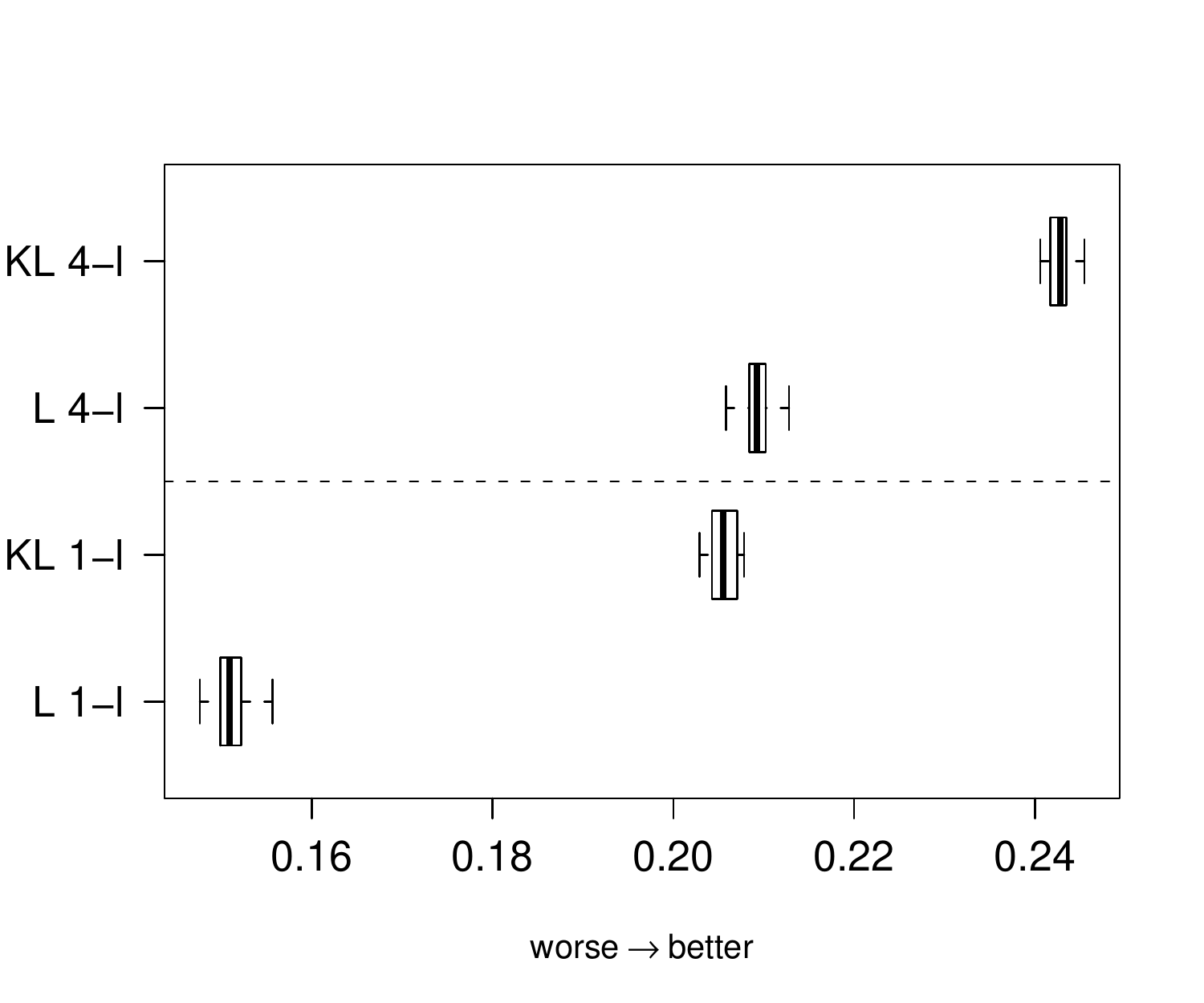}}
 \subfigure[Values of $\beta_{\rho}$]{\includegraphics[trim=0 0 0 50, width=.32\linewidth]{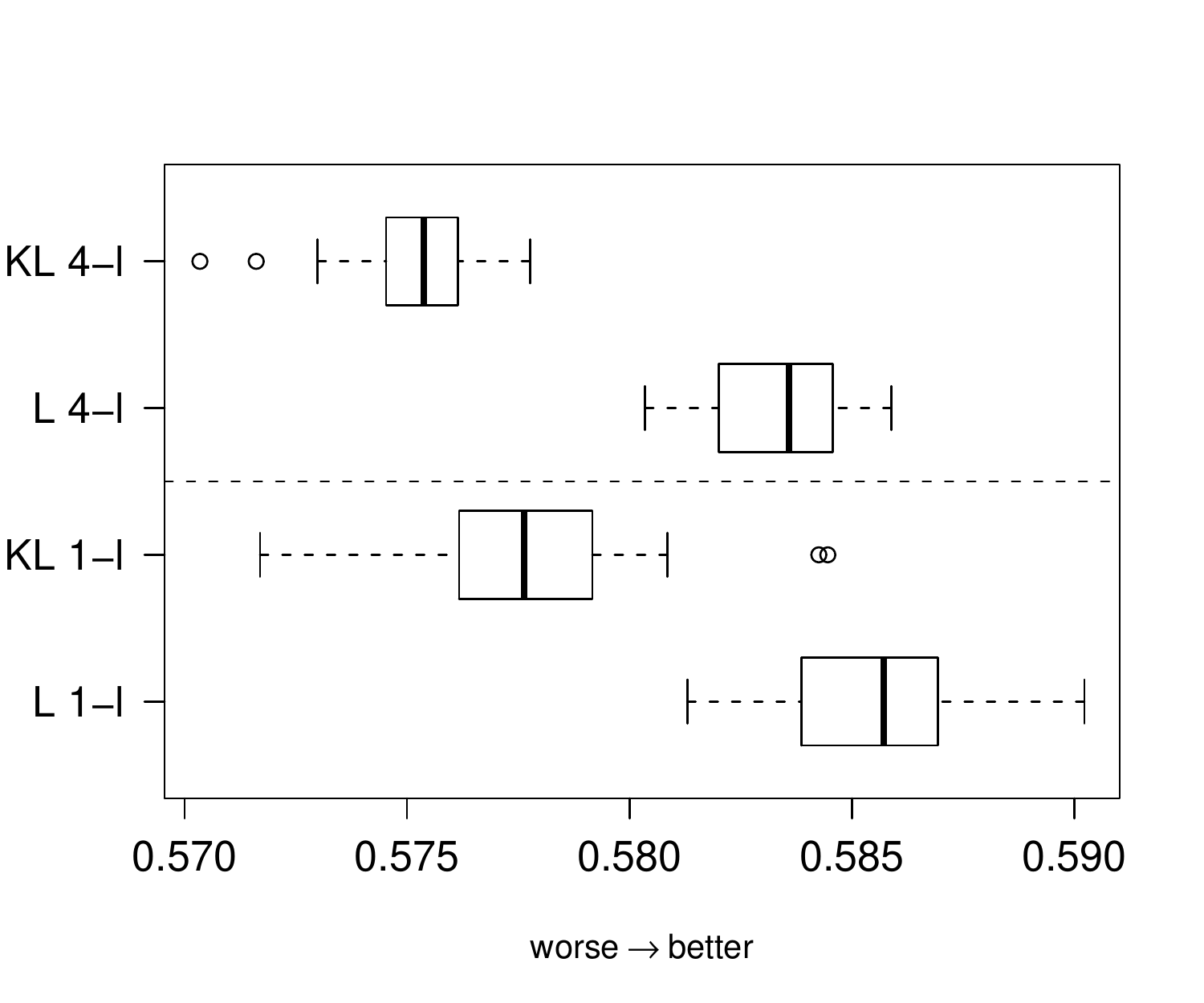}}
 \caption{Boxplots of six metrics applied to four filters.}
 \label{fig:boxplots}
\end{figure*}

The proposal was compared with the Lee filter~\cite{Lee1986} which is considered a standard.
The tests were performed at the $95\%$ level of significance.
The results obtained are summarized by means of boxplots (see Table~\ref{tab:statistics}). 
Each boxplot describes the results of one filter, generating $100$ independent $L=\{1,4\}$ looks images, mean background $\lambda=30$ and mean lines $\lambda=120$.
Figure~\ref{fig:boxplots} shows the boxplots of the six metrics corresponding to four filters. 
Vertical axes are coded by the filter (`L' for Lee and `KL' for Kullback-Leibler) on looks images (`$1$-l' for $1$-look and `$4$-l' for $4$-looks).

\begin{table*}[hbt]
\centering
\renewcommand{\arraystretch}{.5}
\small
\caption{Statistics from image quality indexes.}
\begin{tabular}{c c c c c c c c c c} 
\toprule
  & & \multicolumn{4}{c}{SAR Measures} & \multicolumn{2}{c}{$Q$ index} & \multicolumn{2}{c}{$\beta_{\rho}$ index} \\ \cmidrule(lr{.5em}){3-6} \cmidrule(lr{.5em}){7-8} \cmidrule(lr{.5em}){9-10}
  & Speckle& \multirow{2}{*}{\textsf{NEL}} & Line & Edge & Edge & \multirow{2}{*}{$\bar{Q}$} & \multirow{2}{*}{$s_{Q}$} & \multirow{2}{*}{$\bar{\beta}_{\rho}$} & \multirow{2}{*}{$s_{\beta_{\rho}}$} \\ 
  & Filter &  & Pres. & Grad. & Var. &  &  &  & \\ \midrule
\multirow{2}{*}{$1$-look}
  & Lee & 7.673 & 1.877 & 80.902 & \textbf{1.438} & 0.151 & 0.002 & \textbf{0.586} & 0.002  \\ \cmidrule(lr{.5em}){2-10} 
  & KL  & \textbf{9.305} & \textbf{1.831} & \textbf{69.447} & 4.690 & \textbf{0.206} & 0.002 & 0.578 & 0.003  \\ \midrule
\multirow{2}{*}{$4$-looks}
  & Lee & 14.541 & 1.815 & 59.415 & 7.239 & 0.209 & 0.001 & \textbf{0.583} & 0.002  \\ \cmidrule(lr{.5em}){2-10} 
  & KL  & \textbf{24.442} & \textbf{1.758} & \textbf{47.114} & \textbf{6.83} & \textbf{0.243} & 0.001 & 0.575 & 0.001  \\ 
\bottomrule
\end{tabular} 
\label{tab:statistics}
\end{table*}

The proposal outperforms the Lee filter with respect to equivalent number of looks, line preservation, edge gradient, edge variance on $4$-looks and universal quality index, while the Lee filter presents better performance with respect to the edge variance on $1$-look and the $\beta_{\rho}$ factor.
In all cases the differences are significative.

\section{Conclusions}\label{sec:conclu}

This paper presented a new filter based on stochastic distances for speckle noise reduction.
The proposal was compared with the classical Lee filter, using a protocol based on Monte Carlo experiences. 
Moreover, the $\beta_{\rho}$ and $Q$ index were used to assert the proposal.
The proposed filters behave alike, and they outperform the Lee filter in five out of six quality measures.
Other significance levels will be tested, along with different points of the parameter space in order to have a more complete assessment of the proposal.

\bibliographystyle{IEEEbib}
\bibliography{TorresIGARSS2012}

\end{document}